%% file: EtaEtaPrime.tex
\documentclass[%
superscriptaddress,
preprint,
nofootinbib,
 amsmath,amssymb,
 aps,
]{revtex4-1}

\usepackage[pdftex]{graphicx}% Include figure files
\usepackage{epsfig} % convert eps to pdf
\usepackage{dcolumn}% Align table columns on decimal point
\usepackage{bm}% bold math
\usepackage{hyperref}% add hypertext capabilities
\usepackage{color}% allow colored text
\usepackage[normalem]{ulem}
\usepackage{upgreek}
\begin{document}
%
%\preprint{GlueX-Doc 4093.v8}
%  
\title{Beam Asymmetry $\mathbf{\Sigma}$ for the Photoproduction of $\mathbf{\eta}$ and $\mathbf{\eta^{\prime}}$ Mesons at $\mathbf{E_{\gamma}=8.8}$~GeV}
\date{November 12, 2019}
%
%\author{GlueX Collaboration}
\include{authors}
\begin{abstract}
We report on the measurement of the beam asymmetry $\Sigma$ for the reactions $\vec{\gamma}p\rightarrow p\eta$ and $\vec{\gamma}p \rightarrow p\eta^{\prime}$ from the GlueX experiment, using an 8.2--8.8~GeV linearly polarized tagged photon beam incident on a liquid hydrogen target in Hall D at Jefferson Lab. These measurements are made as a function of momentum transfer $-t$, with significantly higher statistical precision than our earlier $\eta$ measurements, and are the first measurements of $\eta^{\prime}$ in this energy range. We compare the results to theoretical predictions based on $t$--channel quasi-particle exchange. We also compare the ratio of $\Sigma_{\eta}$ to $\Sigma_{\eta^{\prime}}$ to these models, as this ratio is predicted to be sensitive to the amount of $s\bar{s}$ exchange in the production. We find that photoproduction of both $\eta$ and $\eta^{\prime}$ is dominated by natural parity exchange with little dependence on $-t$. 
\end{abstract}
\maketitle

%\section{Introduction}
Photoproduction of $\eta$ and $\eta^{\prime}$ mesons has been important in the search for isospin $\frac{1}{2}$ baryon resonances, with both cross section and spin observables providing input in this endeavor. In the nucleon resonance region, the $s$--channel baryon production is mixed with $t$--channel Reggeon exchange, while at high energy (above 7~GeV), reactions are dominated by the $t$--channel contributions~\cite{Barker:1975bp,Mathieu:2015gxa}. Of particular interest in the high-energy region is the photon beam asymmetry $\Sigma$, measured using linearly polarized photons. This observable is sensitive to the naturality of the exchange particle~\cite{Nys:2016vjz}, and a determination of the beam asymmetries for the $\eta$ and $\eta^{\prime}$ ($\Sigma_{\eta}$ and $\Sigma_{\eta^{\prime}}$, respectively) at high energy directly constrains these same contributions at lower energies. While $\Sigma_{\eta}$ and $\Sigma_{\eta^{\prime}}$ provide valuable information on their own, the ratio of the two can shed light on the contributions of hidden strangeness exchange ($s\bar{s}$ states, such as the $\phi$ and $h_1^\prime$) and axial vector meson ($b$ and $h$) exchange~\cite{Mathieu:2017jjs}.

There is substantial literature of photon beam asymmetry measurements for the $\eta$ below 4~GeV beam energies~\cite{Vartapetian:1980cn,Bussey:1980mz,Ajaka:1998zi,Elsner:2007hm,Bartalini:2007fg,Fantini:2008zz,Collins:2017sgu}. A more limited set of $\Sigma_{\eta^{\prime}}$ measurements exists in the same energy region~\cite{Sandri:2014nqz,Collins:2017sgu}, however, only one measurement of $\Sigma_{\eta}$ above 7~GeV exists~\cite{AlGhoul:2017nbp}.

In this paper, we extend our earlier measurement of the linearly polarized photon beam asymmetry of the $\eta$ meson~\cite{AlGhoul:2017nbp} in $\vec{\gamma} p\rightarrow p\eta$ with more precise measurements. We also report the first measurement of the beam asymmetry of the $\eta^{\prime}$ photoproduction in the photon energy range $8.2-8.8$~GeV (flux-averaged beam energy is $8.5$~GeV). These data have been acquired during the first dedicated physics running of GlueX in Hall~D of the newly upgraded Continuous Electron Beam Accelerator Facility (CEBAF) at Jefferson Lab. They represent an integrated luminosity of $20.8$~pb$^{-1}$ collected at a beam pulse repetition rate of $250$~MHz in GlueX.

%\section{Experimental Details}

Tagged photons are produced through the processes of bremsstrahlung and coherent bremsstrahlung by passing the $11.6$~GeV CEBAF electron beam through an aligned $50\,\upmu$m thick diamond radiator and measuring the energy of each recoil electron using a highly segmented hodoscope, which covers the 8.2--8.8~GeV energy range of the coherent bremsstrahlung peak and allows us to determine each photon's energy with an accuracy of  $\approx10$~MeV. Four orientations of the diamond radiator are used to produce two sets of orthogonal linear polarizations, one set parallel and perpendicular to the lab floor (referred to as `0/90'), and a second set, rotated by $45^{\circ}$ from the first (`-45/45'). About $10$\% of the data have been collected using a $30\,\upmu$m thick aluminum radiator, while the remaining data are equally divided among the four diamond orientations. Data were taken by cycling through each of the five configurations, with about two hours of data collection in each configuration, per cycle.

\begin{figure}[!ht]\centering
\includegraphics[width=0.7\linewidth]{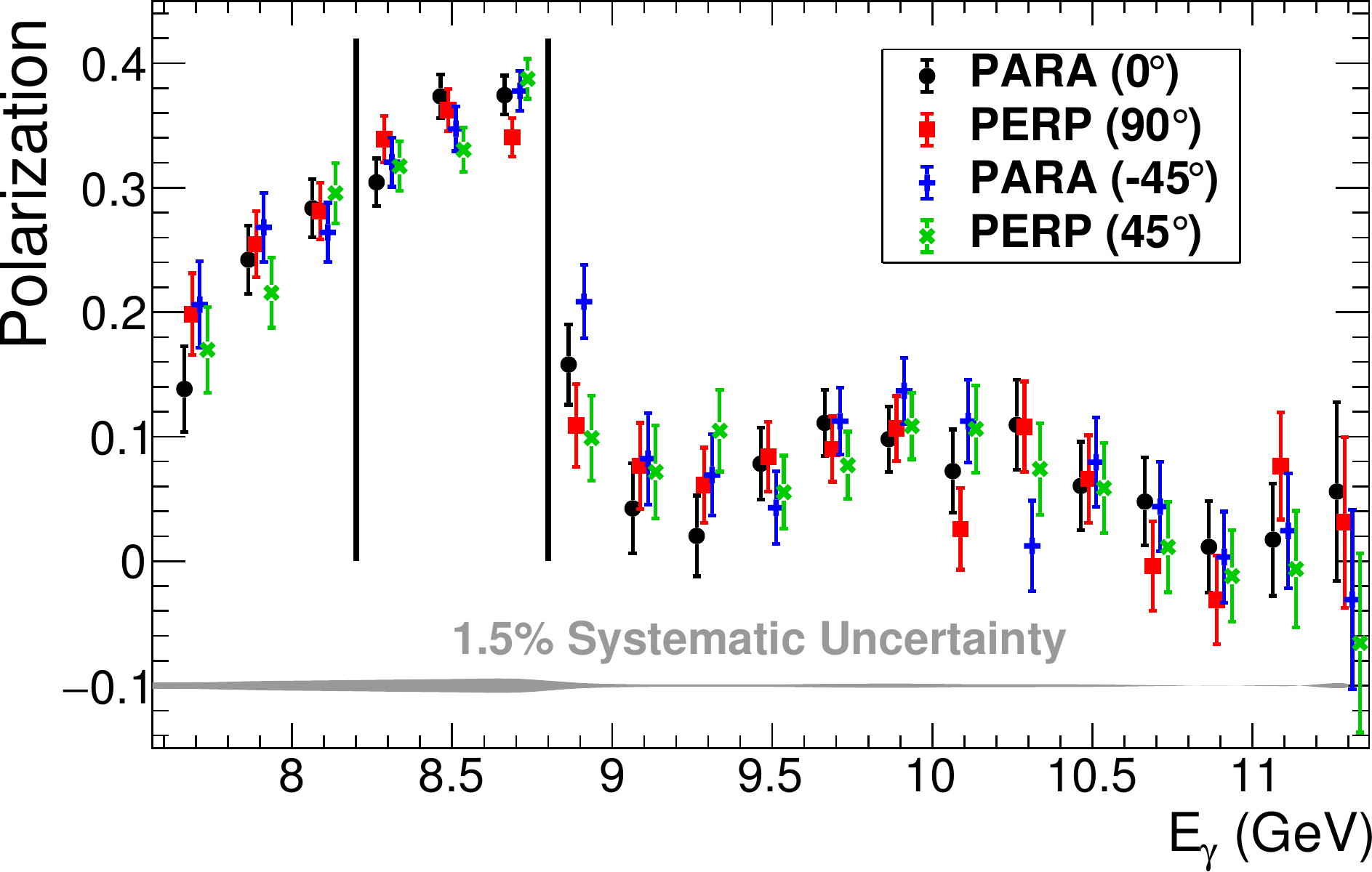}
\caption[]{\label{fig:polarization}The measured degree of linear polarization for the four diamond orientations is plotted as a function of the photon energy, offset from one another in energy for clarity. Events with energy between 8.2--8.8~GeV are selected, as demarcated by the vertical lines.}
\end{figure}

The produced photons travel $75$~m before passing through a $5$~mm diameter collimator, which removes off-axis photons from the beam. This enhances the fraction of coherently produced photons, yielding a photon beam with peak linear polarization of $40$\%, as shown in Fig.~\ref{fig:polarization}. The energy and flux of the photon beam are measured by a pair spectrometer~\cite{Barbosa:2015bga}, which detects pair production of $e^{+}e^{-}$ in a $75\,\upmu$m thick beryllium converter. The polarization of the photons is measured using a triplet polarimeter~\cite{Dugger:2017zoq} using the process $\vec{\gamma}e^{-}\rightarrow e^{-}e^{+}e^{-}$. The high-energy pair is measured in the pair spectrometer, while the low-energy recoil electron is detected in a $1$~mm thick silicon detector. The photon polarization $P_{\gamma}$ is obtained from the azimuthal angular distribution $\phi_{e}$ of the low-energy electron via
\begin{eqnarray}
\frac{d\sigma}{d\phi_{e}} & \propto & \left[ 1 - P_{\gamma}\lambda \cos 2\left(\phi_{e}-\phi_{\gamma}\right)\right]\, ,
\end{eqnarray}
where $\phi_{\gamma}$ is the orientation of the linear polarization and $\lambda$ is the analyzing power, which is fully determined by quantum electrodynamics.
The measured linear polarization, as a function of the photon energy, is shown for each of the four diamond orientations in Fig.~\ref{fig:polarization}. The average polarization in each orientation is determined from the average of measurements in the coherent peak region, weighted by the beam energy distribution for reconstructed $\eta$ or $\eta^{\prime}$ events. The statistical uncertainties of the average polarizations are driven by the yield of triplet production events in the data sample, while the systematic uncertainty in the design and operation of the triplet polarimeter is 1.5\%~\cite{Dugger:2017zoq}. This uncertainty contributes to the overall relative uncertainty of 2.1\% discussed later.

%%%-----------------------------------
%%%-----------------------------------

The GlueX detector is nearly hermetic and azimuthally symmetric, and optimized for a fixed target photoproduction experiment. It is based on a $\sim$4~m long superconducting solenoid magnet that produces a $\sim$2~T field. The solenoidal magnetic field confines low energy electromagnetic background ($e^{+}e^{-}$ pairs) generated in the target to within a small radius of the photon beamline. Inside the bore of the solenoid, the incident photons interact in a $30$~cm long liquid hydrogen target that is located $65$~cm from the upstream end of the solenoid. The target is surrounded by a scintillator-based Start Counter (ST) that records the time of charged particles ~\cite{Pooser:2019rhu}, and a Central Drift Chamber (CDC) that contains $28$ layers of $1.5$~m long, $1.6$~cm diameter straws arranged in axial and stereo orientations~\cite{VanHaarlem:2010yq}. Downstream of the CDC and at forward angles are four planar packages of forward drift chambers (FDC)~\cite{Berdnikov:2015jja,Pentchev:2017omk}. Charged particle tracks are reconstructed with momentum resolution between $1\%$ and $7\%$, depending on their angle and momentum. The drift chambers also provide energy-loss information which allows for $\pi$-$p$ separation up to about $1$~GeV/c momentum. A lead--scintillating-fiber barrel calorimeter (BCAL) encompasses all the drift chambers and measures the position, energy, and time of all incident particles~\cite{Beattie:2018xsk}. Downstream past the solenoid is a scintillator-based time-of-flight (TOF) wall that measures the arrival time of charged particles. A forward calorimeter (FCAL) is located downstream of the TOF wall and measures the energy, position, and time of particles in a $2800$-element array of lead-glass blocks~\cite{Moriya:2013aja}.

The data for this study were reconstructed in two exclusive final states: $\vec{\gamma}p\rightarrow p\gamma\gamma$ for the $\eta$ decaying to $\gamma\gamma$, and $\vec{\gamma}p\rightarrow p\pi^{+}\pi^{-}\gamma\gamma$ for the $\eta^{\prime}$ decaying to $\eta\pi^{+}\pi^{-}$. The final states were selected by choosing events with an associated topology: one positively charged track and two photons for the $\eta$, and two positively and one negatively charged track together with two photons for the $\eta^{\prime}$. Protons are identified using momentum and energy-loss information from the drift chambers in the central region, and through time-of-flight in the forward direction.

Initial event selection requires a primary event vertex inside the GlueX target, no photons near the edges of the calorimeters where shower reconstruction is incomplete, and proton momentum above $250$~MeV/c (to ensure that it can be consistently detected in the drift chambers). The time of the primary interaction is determined by hits in the ST matched to the recoil proton track, and is used to specify which beam bunch of electrons is associated with the event, as the accelerator delivers one bunch of electrons every $4$~ns. Photons associated with the primary interaction are selected using the difference between the bunch's time (provided by the accelerator) and the tagged photon's time, ${\Delta t = |t_\text{photon} - t_\text{bunch}| < 2}$~ns. A separate sample of events with ${6 < \mid\Delta t \mid < 18}$~ns, corresponding to six out-of-time beam bunches (three early and three late), is also selected to account for photons accidentally associated with the primary interaction.

To ensure reaction channel exclusivity, a condition is placed on the square of the missing mass of the event, defined as ${\text{MM}^{2}  =  |p_{\text{in}} - p_{\text{fin}}|^{2}}$, where $p_{\text{in}}$ is the sum of the initial state four-momentum vectors (beam photon and target proton), and $p_{\text{fin}}$ is the sum of the final state four-momentum vectors ($p$ and two $\gamma$s for the $\eta$, and $p$, $\pi^{+}, \pi^{-}$, and two $\gamma$s for the $\eta^{\prime}$). The missing mass squared is required to be consistent with zero, $| \text{MM}^{2} |   \leq   0.05~(\text{GeV}/c^2)^2$, which reduces contributions from massive particles not detected in the event. As an additional condition of exclusivity, both channels excluded events containing extra photons that did not appear to be part of the reconstructed event. 

Next, kinematic fitting is performed on the two exclusive final states. In the case of the $\eta$, a four-constraint fit requiring energy and momentum conservation is performed assuming $\gamma p\rightarrow p\gamma\gamma$. In the case of the $\eta^{\prime}$, an eight-parameter fit is performed for the hypothesis $\gamma p\rightarrow p\pi^{+}\pi^{-}(\eta\rightarrow\gamma\gamma)$, applying energy and momentum conservation and constraining the event vertex and mass of the $\eta$. Selection cuts are placed on the resulting $\chi^{2}$ from the fits to isolate the desired final states. The cut values are the result of detailed studies of the two reactions to optimize signal to background in each channel. Finally, the energy of the beam photon must be in the coherent peak. Detailed Monte Carlo studies of non-exclusive $\eta$ and $\eta^{\prime}$ production processes limit the level of peaking background satisfying all the event selection criteria to less than one part in a thousand. 

\begin{figure}[!ht]\centering
\includegraphics[width=0.65\linewidth]{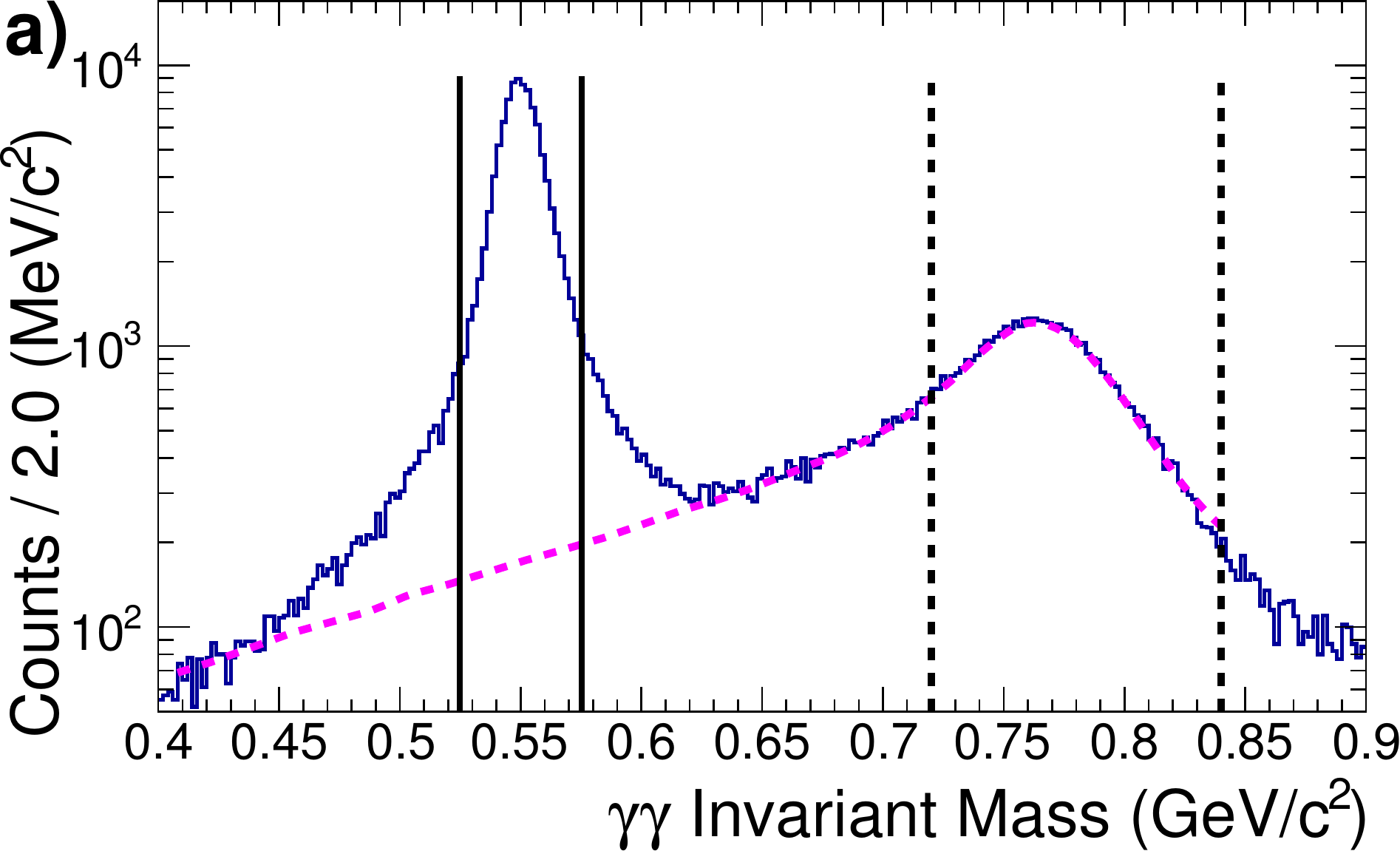} 
\\
  \includegraphics[width=0.65\linewidth]{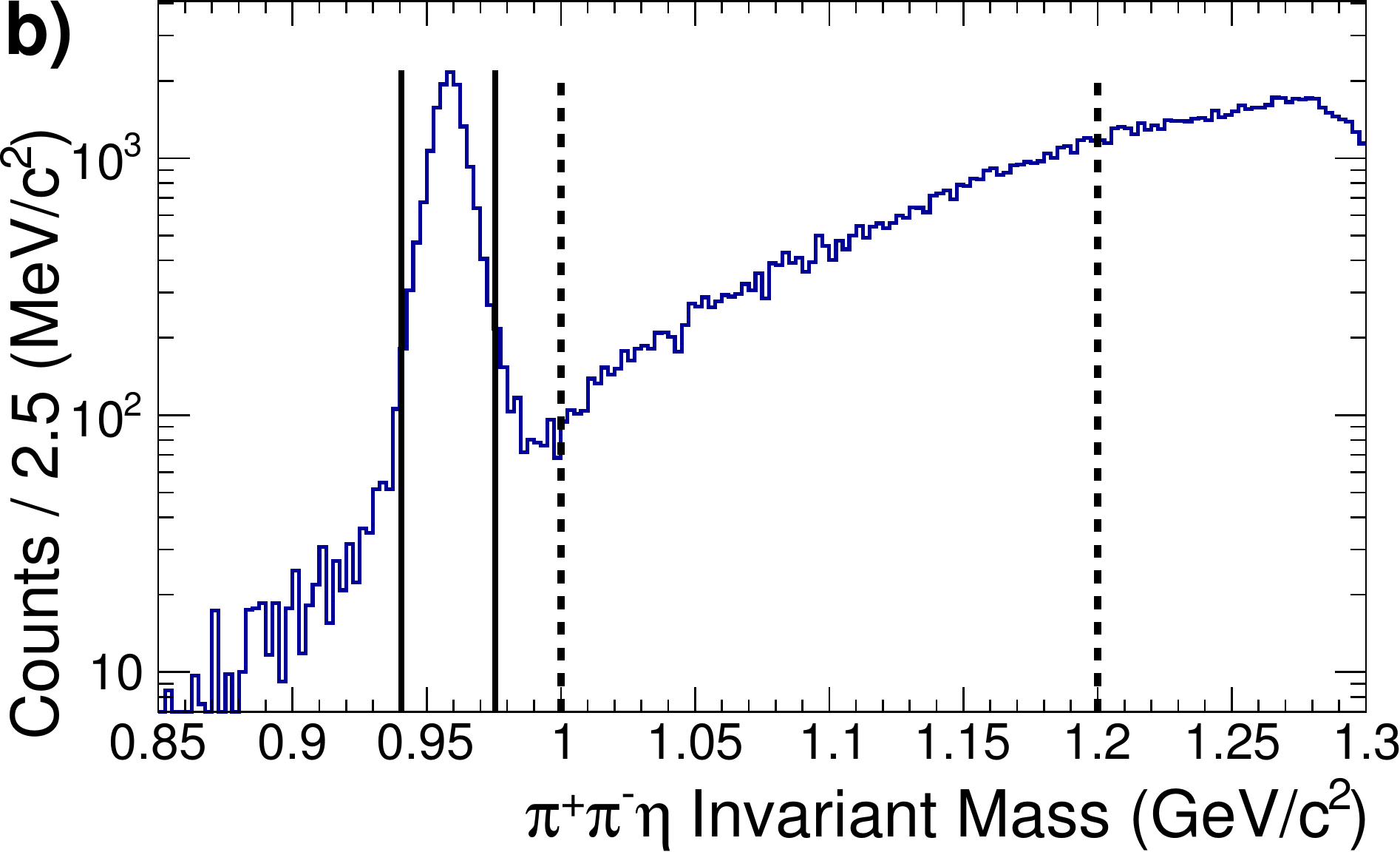}
\caption[]{\label{fig:spectra}The $2\gamma$ (a) and $\pi^+\pi^-\eta$ (b) invariant mass distributions are graphed after all selection cuts are applied. The $\eta$ and $\eta^{\prime}$ `peak region' samples consist of the events between the solid vertical lines. The `side-band region' samples include events between the vertical dashed lines and are used to evaluate the background asymmetry. The dashed curve on (a) is a Monte Carlo calculation of the reaction $\gamma p\rightarrow p\omega$ where the $\omega\rightarrow \pi^{0}\gamma$ and one of the resulting photons is not detected.}
\end{figure}

The same analysis is performed on the out-of-time event sample, and the resulting out-of-time signal is subtracted (with a weight of $\frac{1}{6}$) from the in-time signal. The resulting mass spectra for $\eta$ and $\eta^{\prime}$ candidates are shown in Fig.~\ref{fig:spectra}. Pronounced particle peaks are observed at the expected $\eta$ and $\eta^\prime$ masses, both on top of a small amount of background, described in more detail below. The final event sample is selected by choosing the events between the two vertical lines surrounding the $\eta$ and $\eta^{\prime}$ mass peaks. The treatment of the background contribution to the measured beam asymmetry is discussed later.
%%%-----------------------------------
%%%-----------------------------------

\begin{figure}[!ht]\centering
\includegraphics[width=0.65\linewidth]{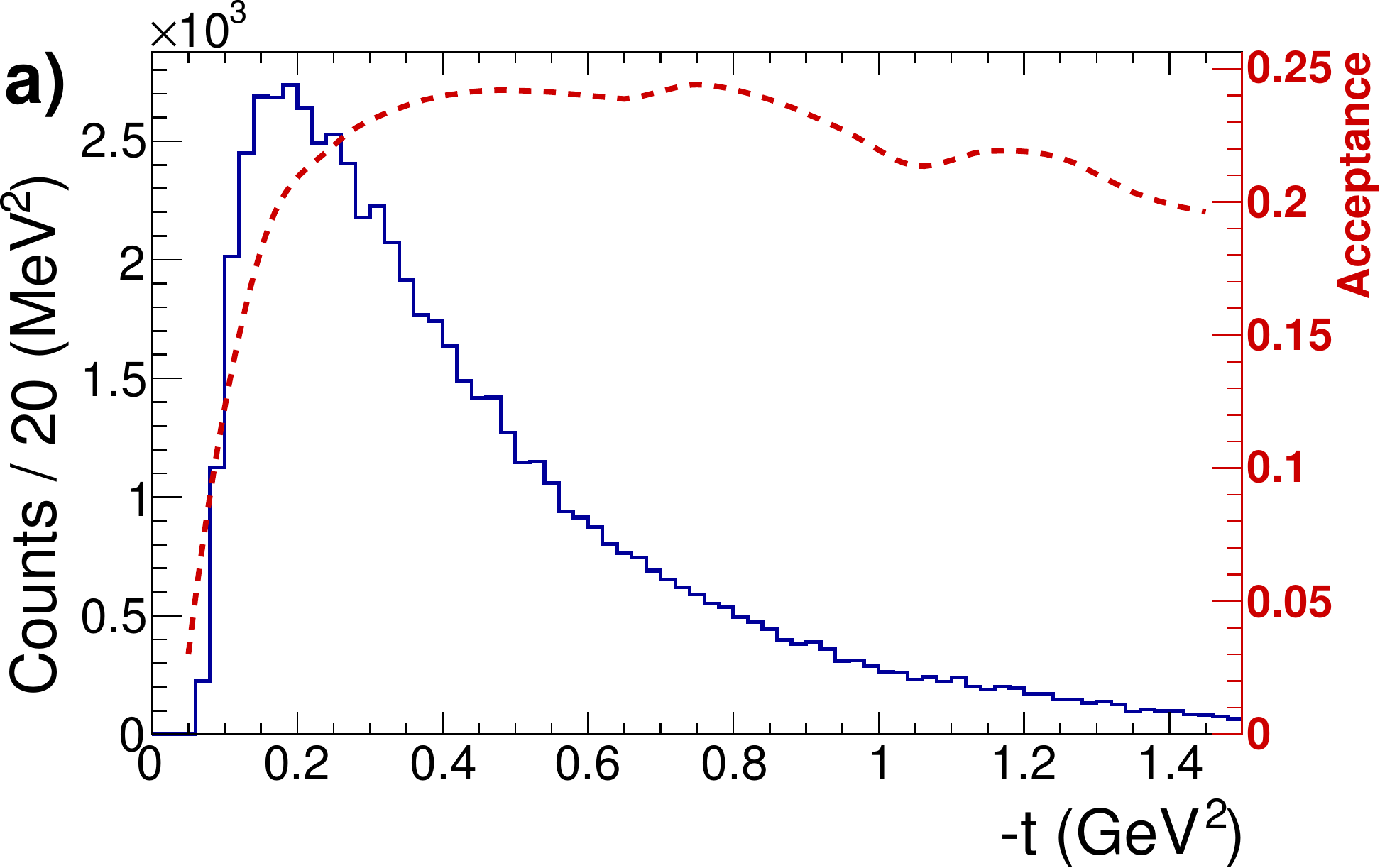}
\\
    \includegraphics[width=0.65\linewidth]{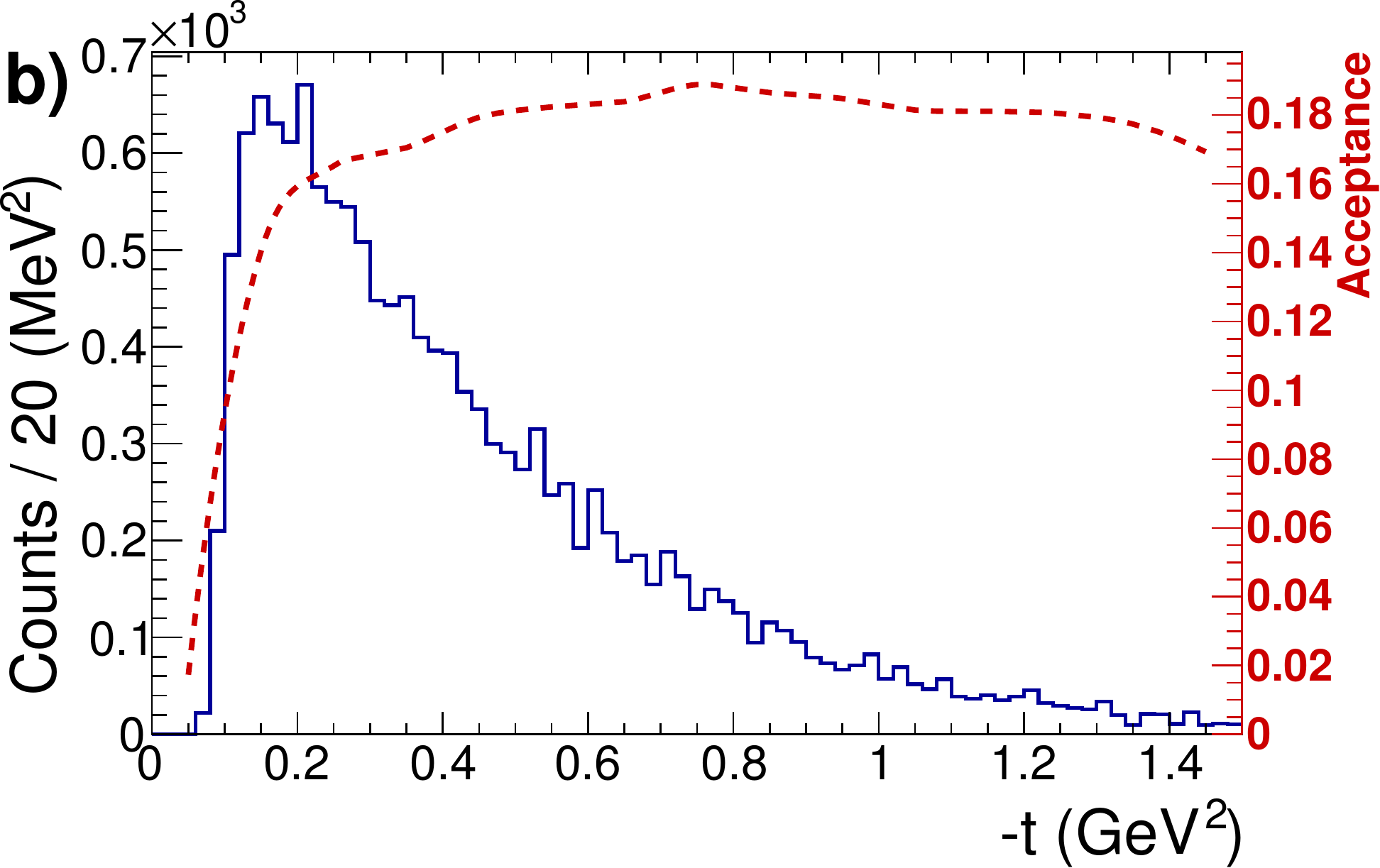}
\caption[]{\label{fig:t}The yields of $\eta$ (a) and $\eta^{\prime}$ (b) events are plotted as a function of $-t$ after all selection cuts are applied. The acceptance functions for $\gamma p\rightarrow\eta p  (p\gamma\gamma)$ and $\gamma p\rightarrow\eta' p  (p\pi^{+}\pi^{-}\gamma\gamma)$, shown as the dashed curves, are determined from Monte Carlo simulation using a Regge model.}
\end{figure}

Using these selection criteria, the yields of $\eta$ and $\eta^{\prime}$ are shown as a function of the momentum transfer $-t$ in Fig.~\ref{fig:t}. The diminishing yield approaching $-t=0.1$~GeV$^{2}$ mainly arises from the $250$~MeV/c cut on the momentum of the recoil proton. The evaluation of the acceptance is based on a Regge model describing the underlying physics in terms of $t$-channel meson exchange and is found to give a reasonable description of the data. Monte Carlo (MC) simulations are performed and compared with data to determine the detector acceptance
as a function of the momentum transfer $-t$ (see Fig.~\ref{fig:t}).
Other than the fall-off at $-t$ near zero, the acceptance is approximately flat, demonstrating that it does not introduce any significant distortion to the yield distributions.
%%%-----------------------------------
%%%-----------------------------------

%\section{\label{sec:Analysis}Analysis}

The analyses are reported in more detail elsewhere~\cite{Beattie:2019tdb,McGingley:2019wm}, while their key steps are summarized herein. For the photoproduction of pseudoscalar mesons with a linearly polarized photon beam and an unpolarized target, the polarized cross section $\sigma_{\mathrm{pol}}$ is related to the beam asymmetry through Eq.~\ref{eq:sigma-pol},
\begin{eqnarray}
\label{eq:sigma-pol}
\sigma_{\mathrm{pol}}(\phi, \phi_{\gamma}) & = & \sigma_0 \left( 1 - P_\gamma\,\Sigma\, \cos\left[2(\phi - \phi_{\gamma})\right]\right)\, ,
\end{eqnarray}
where $\sigma_0$ is the unpolarized cross section, $P_{\gamma}$ is the magnitude of the photon beam polarization, $\phi$ is the azimuthal angle of the production plane, and $\phi_{\gamma}$ is the azimuthal angle of the photon beam's linear polarization plane determined by the orientation of the diamond radiator. In general, the azimuthal ($\phi$) distribution of the event yield is given by
\begin{align}
Y_\parallel(\phi,\phi_{\gamma}=0) &\propto N_\parallel \left[ \sigma_0 A(\phi)  \left(1-P_\parallel \Sigma \cos2\phi \right)  \right] \\ 
Y_\perp(\phi,\phi_{\gamma}=90) &\propto N_\perp \left[ \sigma_0 A(\phi)  \left(1+P_\perp \Sigma \cos2\phi \right) \right],
\end{align}
\noindent where $A(\phi)$ is an arbitrary function for the $\phi$-dependent detector acceptance and efficiency, and $N_{\perp(\parallel)}$ is the flux of photons in two orthogonal orientations.

The GlueX detector is designed to be symmetric in $\phi$ and thus have a uniform acceptance and efficiency, but here we consider the general case of an arbitrary $\phi$-dependent detector acceptance and define the method for extracting $\Sigma$ that cancels this detector acceptance. We choose the diamond radiator orientation such that we have two sets of orthogonally polarized data, which causes the detector acceptance effects to cancel when forming the \emph{yield asymmetry}, as in Eq.~\ref{eqn:asym}:
\begin{equation}
\frac{Y_{\perp}(\phi)\!-\!F_{R} Y_{\parallel}(\phi)}{Y_\perp(\phi)\!+\!F_{R} Y_\parallel(\phi)} = \frac{(P_{\perp}\!+\!P_{\parallel}) \Sigma \cos2\left(\phi\! -\!\phi_{0}\right)}{2\!+\!(P_{\perp}\!-\!P_{\parallel}) \Sigma \cos2\left(\phi\!-\!\phi_{0}\right)}\, .
\label{eqn:asym}
\end{equation}
In this equation we introduce the \emph{phase offset} $\phi_{0}$ which accounts for slight misalignment in the orientation of the polarization plane ($\phi_{\gamma}$) away from its nominal value. The value os $\phi_{0}$ is found to be small (about $3^{\circ}$).

\begin{figure}[!ht]\centering
  \includegraphics[width=0.65\linewidth]{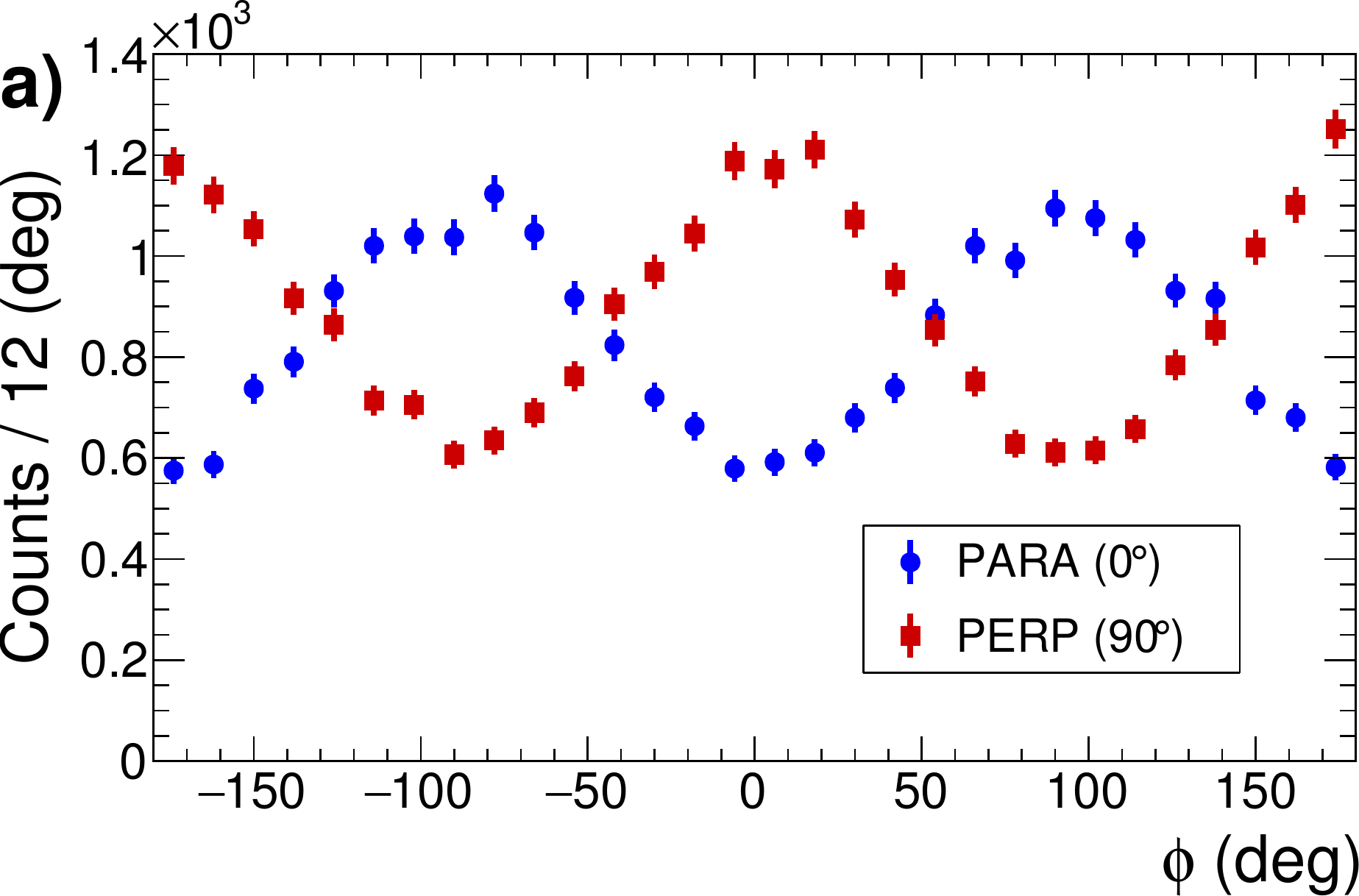}
\\
\includegraphics[width=0.65\linewidth]{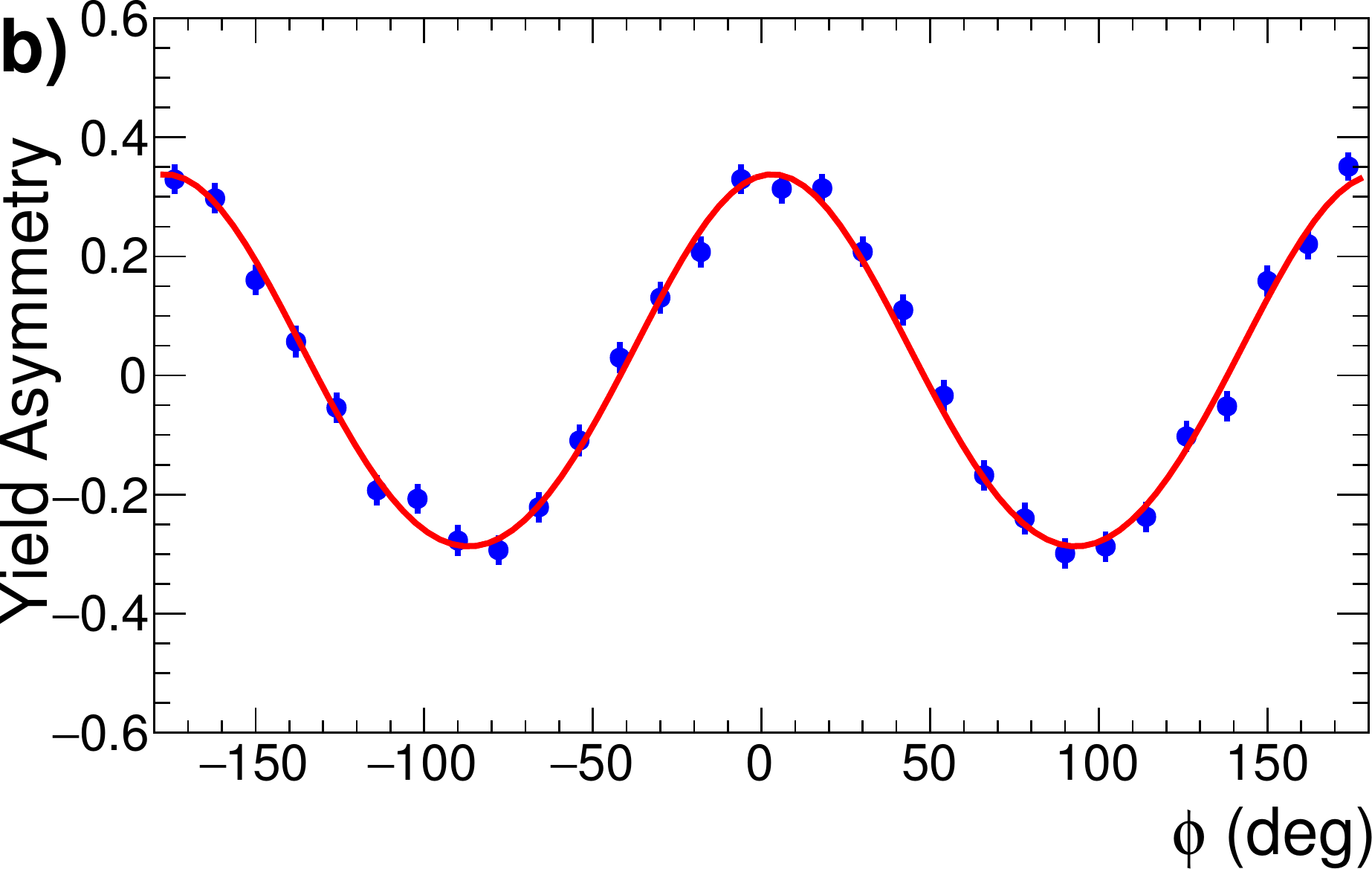}
  \caption[]{\label{fig:Example-Fit} (a) The yields integrated over the full range of $-t$, $Y_{\perp}$ and $Y_{\parallel}$, are shown for the $\eta$ events using one set of orthogonally polarized data, and (b) the yield asymmetry is shown, fitted with a $\chi^2/\mathrm{ndf} = 25.59/28$.}
\end{figure}

The flux normalization ratio $F_R = \frac{N_\perp}{N_\parallel}$ is the ratio of the integrated photon flux for the two orthogonal orientations of the photon polarization. For the 0/90 set, ${F_R = 1.038 \pm 0.052}$, while for the -45/45 set, ${F_R = 0.995 \pm 0.050}$.  The yield asymmetry is formed for the $\eta$ and $\eta^{\prime}$ in bins of $-t$, and $\Sigma$ is extracted in each bin through fits of  Eq.~\ref{eqn:asym} to the asymmetry data, where $\Sigma$ is the only free parameter. Fig.~\ref{fig:Example-Fit}(a) shows the yields, $Y_{\perp}$ and $Y_{\parallel}$, for the $\eta$ events (integrated over all values of $-t$) as a function of the angle $\phi$. The oscillations of the two polarization orientations are $90^{\circ}$ out of phase. Fig.~\ref{fig:Example-Fit}(b) shows the yield asymmetry given by Eq.~\ref{eqn:asym} and the resulting fit to the data.

%%%-----------------------------------
%%%-----------------------------------

In order to correct for possible asymmetries from background events under the $\eta$ and $\eta^{\prime}$ events, the same asymmetry analysis is carried out for background events in the side-band regions as shown in Fig.~\ref{fig:spectra}. The \emph{side-band asymmetry} $\Sigma_\text{SB}$ and the \emph{dilution factor} $f$ (the fractional background under the peak) are extracted. The corrected beam asymmetry $\Sigma_\text{COR}$ is then given by Eq.~\ref{eqn:correctedasym}:

\begin{eqnarray}
\label{eqn:correctedasym}
\Sigma_\text{COR} & = & \frac{\Sigma_\text{peak} - f \Sigma_\text{SB}}{1 - f}
\end{eqnarray}
where $\Sigma_\text{peak}$ is the asymmetry measured in the peak region. This correction shifts the asymmetry values by a few percent in the lowest $-t$ bin, falling to a negligible amount at large $-t$.

Fits to the invariant mass spectra (Fig.~\ref{fig:spectra}) are carried out to extract $f$ for each bin of $-t$. $\Sigma_\text{SB}$ is estimated from a fit of Eq.~\ref{eqn:asym} to the yield asymmetry in the side-band region data. The binning in $-t$ is optimized so that each bin contains an approximately equal number of events; the higher statistics $\eta$ channel allows finer binning than the $\eta^\prime$ channel. Since the background under the $\eta$ peak is almost entirely due to ${\omega \rightarrow \pi^0\gamma}$ events with a missing photon~\cite{AlGhoul:2017nbp} (as shown in Fig.~\ref{fig:spectra}a), the $\Sigma$ asymmetry for background events under the $\eta$ peak is assumed to be identical to the $\Sigma$ asymmetry of events in the $\omega$ peak. Therefore, the side-band region chosen to determine the $\eta$ background asymmetry, ${0.72 < M_{2\gamma} < 0.84}$~GeV/$c^2$, encompasses the $\omega$ peak.  A systematic uncertainty on $\Sigma_\eta$, associated with the $\Sigma_{\text{SB}}$ correction, is assigned to each $-t$ bin and is between $0.2$--$0.4\%$. The background under the $\eta^\prime$ peak comes from multiple, higher lying channels, and the measured asymmetry in the side-band region is mass-dependent. Thus, the assumption that the asymmetry in a mass side-band region is the same as the asymmetry of the background events under the peak may not be completely valid.  However, due to low statistics at high $-t$, a wide mass range is used for the side-band region, ${1.0 < M_{\pi^+\pi^-\eta} < 1.2}$~GeV/$c^2$. With this wide range, mass-dependent effects to the asymmetry are encapsulated in a systematic uncertainty on $\Sigma_{\eta^\prime}$ for each $-t$ bin, between $0.6$--$1.6\%$.

The measured beam asymmetries contain additional sources of systematic uncertainties that are estimated for each of the reported $-t$ bins and are tabulated in Tab.~\ref{tab:systematics}. When the uncertainty varies between $-t$ bins, a range is reported. The largest of these systematic uncertainties is associated with the event selection, and is found by evaluating the asymmetries in each $-t$ bin under varied selection criteria. The errors on the flux normalization ratios, $F_R$, manifest as systematic uncertainties on the $\eta$ and $\eta^\prime$ asymmetries, and, finally, there is an uncertainty associated with the phase offset, $\phi_{0}$. None of these systematic uncertainties are correlated, so they are added in quadrature to give the total systematic uncertainty. In addition to the systematic uncertainties in the analysis, there is a $2.1\%$ relative uncertainty associated with the photon beam polarization that would result in an overall shift in the measured beam asymmetries. We do not combine this with the other uncertainties.
\begin{table}[!ht]\centering
\caption{\label{tab:systematics} Summary of the systematic uncertainties assigned to $\Sigma_\eta$ and $\Sigma_{\eta^\prime}$. See the text for details.}
\begin{ruledtabular}
\begin{tabular}{ccc}
 & \multicolumn{2}{c}{Uncertainties} \\ \cline{2-3}
Source & $\Sigma_\eta$ & $\Sigma_{\eta^\prime}$ \\ \hline
Event Selection & $1.6$--$3.5\%$ & $3.5$--$7.5\%$ \\
$\Sigma_{\text{SB}}$ Correction & $0.2$--$0.4\%$ & $0.6$--$1.6\%$ \\ 
Flux Normalization & $0.2\%$ & $0.4\%$ \\
Phase Offset & $0.1\%$ & $0.5\%$ \\ \hline
Total Systematic Error & $1.6$--$3.5\%$ & $3.7$--$7.6\%$
\end{tabular}
\end{ruledtabular}
\end{table}

%\section{\label{sec:results}Results}

\begin{figure}[!ht]\centering
\includegraphics[width=0.70\linewidth]{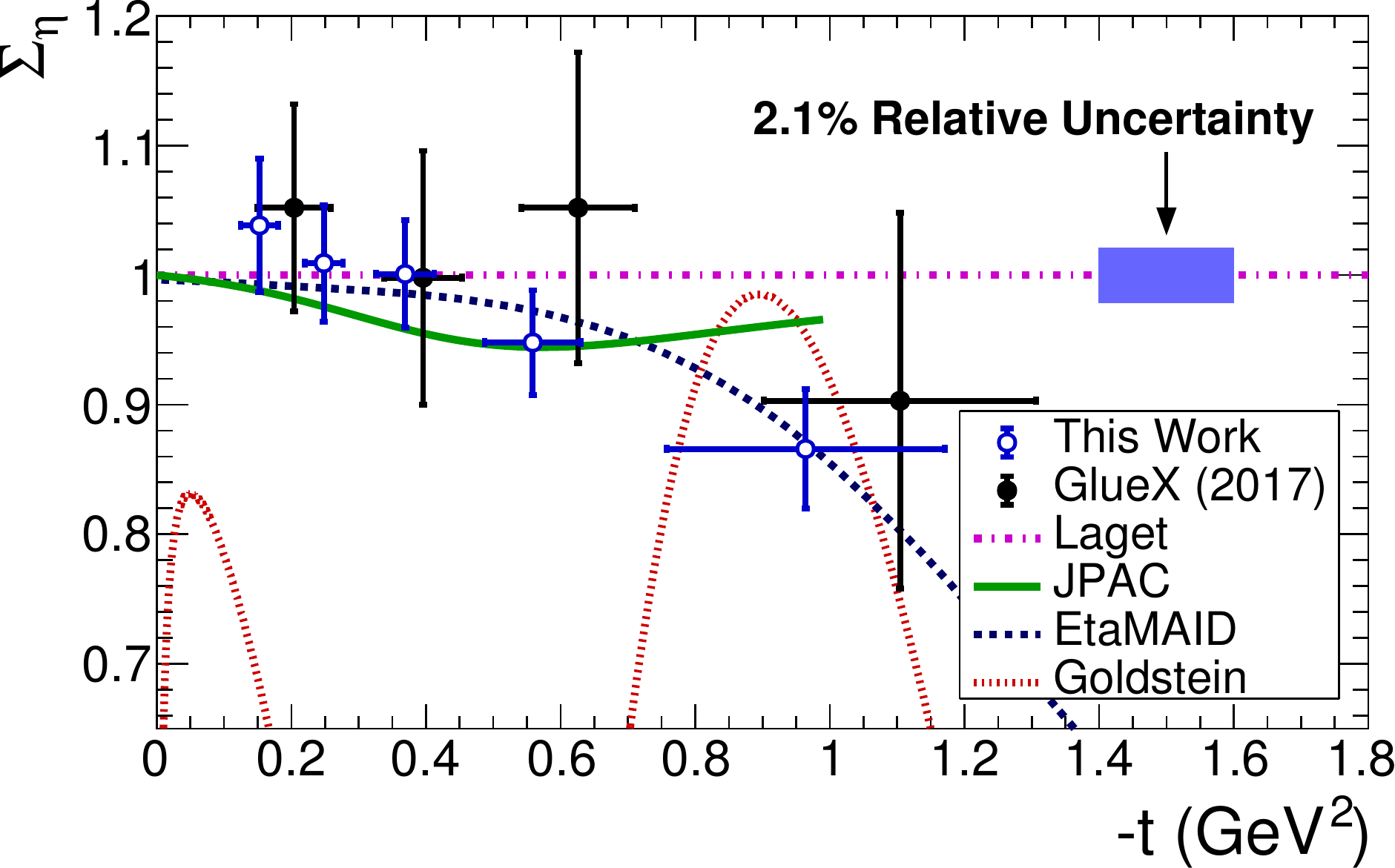}
 \caption[]{\label{fig:etacomparison}The photon beam asymmetry $\Sigma_{\eta}$  is shown as a function of $-t$ for ${\vec{\gamma}p \rightarrow p\eta}$. The vertical error bars represent the total errors and the horizontal error bars represent the RMS widths of the $-t$ distributions in each bin. Previous GlueX (2017) results~\cite{AlGhoul:2017nbp} are shown along with predictions from several Regge theory calculations: Laget~\cite{Laget:2005,Laget:2011}, JPAC~\cite{Nys:2016vjz}, EtaMAID~\cite{Kashevarov:2017vyl} and Goldstein~\cite{Goldstein:1973}. The 2.1\% relative uncertainty is due largely to the polarization measurement.}
\end{figure}

The final photon beam asymmetry results are the weighted averages of the two independent polarization data sets plotted as functions of $-t$. The results for $\Sigma_{\eta}$ are shown in Fig.~\ref{fig:etacomparison}, where they are compared to earlier GlueX data~\cite{AlGhoul:2017nbp} as well as several theoretical predictions (values can be found in Ref.~\cite{supplemental}). For values of $-t$ below $0.6\,(\text{GeV}/c)^2$, the Laget~\cite{Laget:2005,Laget:2011},  JPAC~\cite{Nys:2016vjz} and EtaMAID~\cite{Kashevarov:2017vyl} models describe the data. For $-t$ larger than $0.6\,(\text{GeV}/c)^2$, the Laget and JPAC models appear to overestimate $\Sigma_{\eta}$, while the EtaMAID is in better agreement with the data which suggests that the beam asymmetry may be decreasing with increasing $-t$. The older model by Goldstein~\cite{Goldstein:1973} predicts a lower value of $\Sigma_{\eta}$ than is observed, as well as significant structure, which is not observed. In terms of the models, values of $\Sigma_{\eta}$ near one indicate the reaction is dominated by natural parity exchange mechanisms, while values below one suggest a contribution from unnatural parity exchange as well.
%%%-----------------------------------
%%%-----------------------------------

\begin{figure}[!ht]\centering
  \includegraphics[width=0.70\linewidth]{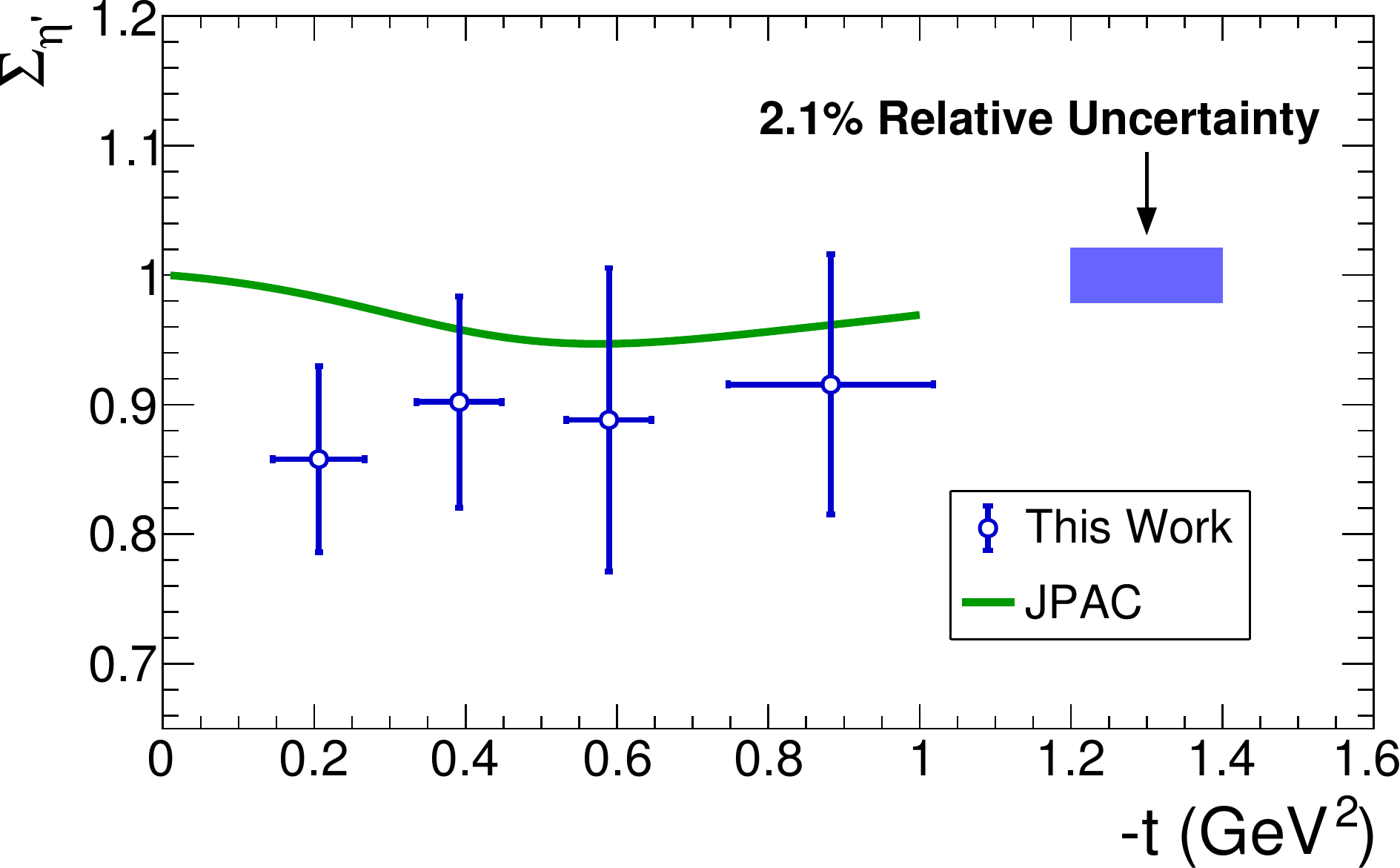}
  \caption{\label{fig:etaprimecomparison} The photon beam asymmetry $\Sigma_{\eta\prime}$ is shown for ${\vec{\gamma}p \rightarrow p\eta^\prime}$. The vertical error bars represent the total errors and the horizontal error bars represent the RMS widths of the $-t$ distributions in each bin. The Regge theory calculation from JPAC~\cite{Mathieu:2017jjs} is shown. The 2.1\% relative uncertainty is due largely to the polarization measurement.}
\end{figure}

The photon beam asymmetry  $\Sigma_{\eta\prime}$  is shown as a function of $-t$ in Fig.~\ref{fig:etaprimecomparison} (values can be found in Ref.~\cite{supplemental}). The results are systematically smaller than one, averaging at around $0.9$ over all values of $-t$. This indicates that while the production of $\eta^{\prime}$ is dominated by natural parity exchanges, there must be some unnatural parity exchange contributions as well. The only theoretical prediction, from JPAC~\cite{Mathieu:2017jjs}, is consistent with these results, but appears to be systematically high. 
%%%-----------------------------------
%%%-----------------------------------

\begin{figure}[!ht]\centering
\includegraphics[width=0.70\linewidth]{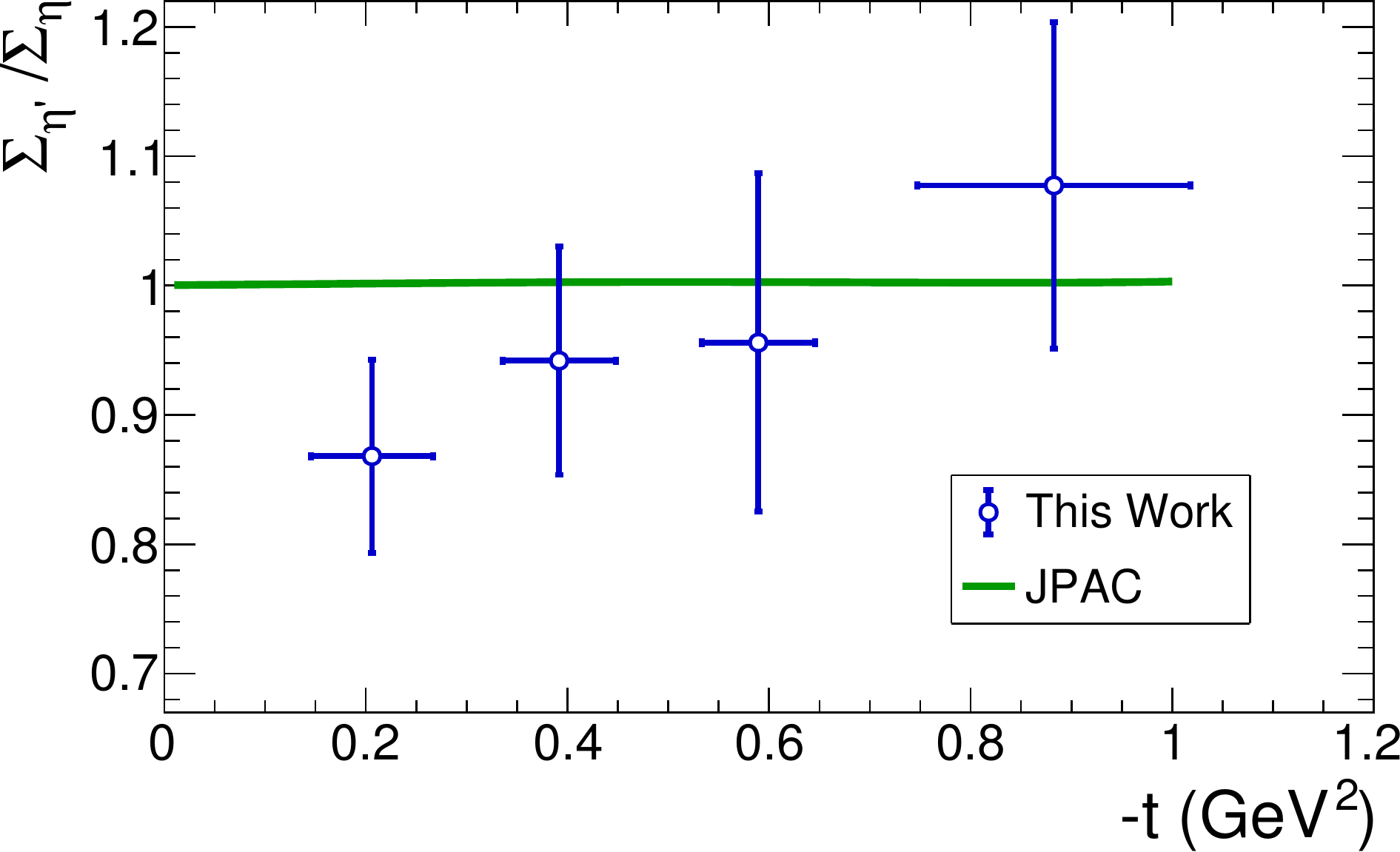}
 \caption {The photon beam asymmetry ratio $\Sigma_{\eta^\prime}/\Sigma_\eta$ is plotted.  The vertical error bars represent total errors. The horizontal error bars represent the RMS widths of the $-t$ distributions in each bin. The Regge theory calculation from JPAC~\cite{Mathieu:2017jjs} is shown.} 
 \label{fig:SigmaRatio}
\end{figure}

In addition to $\Sigma_{\eta\prime}$, the JPAC model~\cite{Mathieu:2017jjs} also predicts the ratio of the beam asymmetries, $\Sigma_{\eta^\prime}/\Sigma_\eta$. We show this ratio in Fig.~\ref{fig:SigmaRatio}, along with the JPAC prediction (values can be found in Ref.~\cite{supplemental}). Because of strong correlations between systematic uncertainties in the two channels, we estimate the systematic on the ratio as the uncorrelated part of the $\eta^{\prime}$ systematic uncertainty. In the JPAC model, a deviation of the ratio from one or even a slope in the distribution suggests that $s\bar{s}$ exchanges ($\phi$ and $h^{\prime}_{1}$) are important in the production. As the measured ratio is consistent with unity, the reactions proceed predominantly through $\rho$ and $\omega$ vector meson exchange. At this time, however, our data are not sensitive enough to be able to draw more detailed conclusions.
%%%-----------------------------------
%%%-----------------------------------
%\clearpage
%\section{\label{sec:summary}Summary}

We have measured the photon beam asymmetry $\Sigma$ for both $\eta$ and $\eta^{\prime}$ photoproduction in the GlueX experiment using an 8.2--8.8~GeV linearly polarized tagged photon beam. These measurements were made as a function of momentum transfer $-t$ and, in the case of the $\eta$, are of significantly greater precision than our earlier $\eta$ measurements~\cite{AlGhoul:2017nbp}. For the $\eta^{\prime}$, these represent the first  measurements of $\Sigma_{\eta\prime}$ in this energy range. The beam asymmetries and their ratio are compared to theoretical predictions based on  $t$--channel quasi-particle exchange. The data show that the asymmetries and ratio are close to unity, which implies that the reactions proceed primarily through $\rho$ and $\omega$ vector meson exchange.
%
%\section{Acknowledgements}
The analysis in this article was supported by the Natural Sciences and Engineering Research Council of Canada grant SAPPJ-2018-00021 and by the U.S. Department of Energy, Office of Science, Office of Nuclear Physics under contract DOE Grant No. DE-FG02-87ER40315. We would like to acknowledge the outstanding efforts of the staff of the Accelerator and the Physics Divisions at Jefferson Lab that made the experiment possible. This work was also supported in part by the U.S. Department of Energy, the U.S. National Science Foundation, the German Research Foundation, GSI Helmholtzzentrum f\"{u}r Schwerionenforschung GmbH, the Russian Foundation for Basic Research, the UK Science and Technology Facilities Council, the Chilean Comisi\'{o}n Nacional de Investigaci\'{o}n Cient\'{i}fica y Tecnol\'{o}gica, the National Natural Science Foundation of China, and the China Scholarship Council. This material is based upon work supported by the U.S. Department of Energy, Office of Science, Office of Nuclear Physics under contract DE-AC05-06OR23177. 
%%%%%%%%%%%%%%%%%%%%%%%%%%%
\bibliography{EtaEtaPrime}
%%%%%%%%%%%%%%%%%%%%%%%%%%%
%\clearpage
%\include{supplemental}
%%%%%%%%%%%%%%%%%%%%%%%%%%%
\end{document}

%% file: authors.tex
\affiliation{Arizona State University, Tempe, Arizona 85287, USA}
\affiliation{National and Kapodistrian University of Athens, 15771 Athens, Greece}
\affiliation{Carnegie Mellon University, Pittsburgh, Pennsylvania 15213, USA}
\affiliation{The Catholic University of America, Washington, D.C. 20064, USA}
\affiliation{University of Connecticut, Storrs, Connecticut 06269, USA}
\affiliation{Florida International University, Miami, Florida 33199, USA}
\affiliation{Florida State University, Tallahassee, Florida 32306, USA}
\affiliation{The George Washington University, Washington, D.C. 20052, USA}
\affiliation{University of Glasgow, Glasgow G12 8QQ, United Kingdom}
\affiliation{GSI Helmholtzzentrum f\"ur Schwerionenforschung GmbH, D-64291 Darmstadt, Germany}
\affiliation{Institute of High Energy Physics, Beijing 100049, People's Republic of China}
\affiliation{Indiana University, Bloomington, Indiana 47405, USA}
\affiliation{Alikhanov Institute for Theoretical and Experimental Physics NRC Kurchatov Institute, Moscow, 117218, Russia}
\affiliation{Thomas Jefferson National Accelerator Facility, Newport News, Virginia 23606, USA}
\affiliation{University of Massachusetts, Amherst, Massachusetts 01003, USA}
\affiliation{Massachusetts Institute of Technology, Cambridge, Massachusetts 02139, USA}
\affiliation{National Research Nuclear University Moscow Engineering Physics Institute, Moscow 115409, Russia}
\affiliation{Norfolk State University, Norfolk, Virginia 23504, USA}
\affiliation{North Carolina A\&T State University, Greensboro, North Carolina 27411, USA}
\affiliation{University of North Carolina at Wilmington, Wilmington, North Carolina 28403, USA}
\affiliation{Northwestern University, Evanston, Illinois 60208, USA}
\affiliation{Old Dominion University, Norfolk, Virginia 23529, USA}
\affiliation{University of Regina, Regina, Saskatchewan, Canada S4S 0A2}
\affiliation{Universidad T\'ecnica Federico Santa Mar\'ia, Casilla 110-V Valpara\'iso, Chile}
\affiliation{Tomsk State University, 634050 Tomsk, Russia}
\affiliation{Tomsk Polytechnic University, 634050 Tomsk, Russia}
\affiliation{A. I. Alikhanian National Science Laboratory (Yerevan Physics Institute), 0036 Yerevan, Armenia}
\affiliation{College of William and Mary, Williamsburg, Virginia 23185, USA}
\affiliation{Wuhan University, Wuhan, Hubei 430072, People's Republic of China}
\author{S.~Adhikari}
\affiliation{Old Dominion University, Norfolk, Virginia 23529, USA}
\author{A.~Ali}
\affiliation{GSI Helmholtzzentrum f\"ur Schwerionenforschung GmbH, D-64291 Darmstadt, Germany}
\author{M.~Amaryan}
\affiliation{Old Dominion University, Norfolk, Virginia 23529, USA}
\author{A.~Austregesilo}
\affiliation{Carnegie Mellon University, Pittsburgh, Pennsylvania 15213, USA}
\author{F.~Barbosa}
\affiliation{Thomas Jefferson National Accelerator Facility, Newport News, Virginia 23606, USA}
\author{J.~Barlow}
\author{A.~Barnes}
\affiliation{Carnegie Mellon University, Pittsburgh, Pennsylvania 15213, USA}
\affiliation{University of Connecticut, Storrs, Connecticut 06269, USA}
\author{E.~Barriga}
\affiliation{Florida State University, Tallahassee, Florida 32306, USA}
\author{R.~Barsotti}
\affiliation{Indiana University, Bloomington, Indiana 47405, USA}
\author{T.~D.~Beattie}
\affiliation{University of Regina, Regina, Saskatchewan, Canada S4S 0A2}
% ...
\collaboration{The \textsc{GlueX} Collaboration}